\documentclass[onefignum,onetabnum,a4paper,twocolumn]{article}
\usepackage{epsfig}
\usepackage{epstopdf}
\usepackage{amsmath,amssymb}
\usepackage{textcomp}
\usepackage{gensymb}

\title{Bifurcation analysis of a North Atlantic Ocean box model with two deep-water formation sites} 

\author{{\sc Alannah Neff$^{1}$, Andrew Keane$^{1,2}$\footnote{corresponding author:  {\tt andrew.keane@ucc.ie}}, Henk A. Dijkstra$^{3,4}$ and Bernd Krauskopf$^{5}$} \\ 
$^{1}$School of Mathematical Sciences, University College Cork, Cork, Ireland\\
$^{2}$Environmental Research Institute, University College Cork, Cork, Ireland\\
$^{3}$Institute for Marine and Atmospheric research Utrecht, Department of Physics\\Utrecht University, Utrecht, The Netherlands\\
$^{4}$Center for Complex Systems Studies, Department of Physics, Utrecht University\\Utrecht, The Netherlands\\
$^{5}$Department of Mathematics, University of Auckland, Private Bag 92019\\Auckland, 1142, New Zealand\\
}

\date{May 2023}

\begin{document}

\maketitle

\begin{abstract}

The tipping of the Atlantic Meridional Overturning Circulation (AMOC) to a `shutdown' state due to changes in the freshwater forcing of the ocean is of particular interest and concern due to its widespread ramifications, including a dramatic climatic shift for much of Europe. 
A clear understanding of how such a shutdown would unfold requires analyses of models from across the complexity spectrum. For example, detailed simulations of sophisticated Earth System Models have identified scenarios in which deep-water formation first ceases in the Labrador Sea before ceasing in the Nordic Seas, en route to a complete circulation shutdown.
Here, we study a simple ocean box model with two polar boxes designed to represent deep-water formation at these two distinct sites.
A bifurcation analysis reveals how, depending on the differences of freshwater and thermal forcing between the two polar boxes, transitions to `partial shutdown' states are possible. 
Our results shed light on the nature of the tipping of AMOC and clarify dynamical features observed in more sophisticated models. 

\end{abstract}

%^\pacs{}% insert suggested PACS numbers in braces on next line

%\maketitle %\maketitle must follow title, authors, abstract and \pacs


\section{Introduction}

The Atlantic Meridional Overturning Circulation (AMOC) is the primary circulation pattern in the Atlantic Ocean \cite{dijkstra05}. 
The flow of water is mostly density-driven; that is, water flows from a region of higher density to that of lower density. As a result, in the modern Atlantic Ocean, deep ocean currents transport water from higher to lower latitudes. Simultaneously, warm surface water is carried from the lower latitudes northwards. Eventually, this surface water cools, becomes denser, and sinks to deeper parts of the ocean (so-called \emph{deep-water formation}), and is then transported southwards back towards the lower latitudes. The main sites for deep-water formation in the northern Atlantic Ocean are the Nordic Seas and the Labrador Sea. Altogether, this mass transport of water constitutes part of a thermohaline circulation pattern that transports water across the entire globe. Throughout its history the Earth's global ocean may have experienced many different thermohaline circulation patterns \cite{pohl22}, and it is plausible that this pattern will change again in the future.

Because of the associated heat transport, the AMOC plays a crucial role in moderating the climate of the North Atlantic \cite{bryden01,buckley16} and, as such, it is a widely studied phenomenon. Previous studies have shown that the AMOC is sensitive to variations in freshwater forcing and the mean ocean temperature in the regions of deep-water formation (for example \cite{LEN09,hansen16,brown16,jackson18}) and hence to atmospheric CO$_2$.  
% add more references
In fact, should any of these quantities exceed a certain threshold, the system will suddenly transition to a qualitatively different state, in which the entire AMOC becomes considerably weaker, or completely shuts down. In the literature, this event is referred to as a \emph{critical transition}, or sometimes \emph{tipping event}. A tipping of AMOC would cease the supply of warm surface water, which moderates the climate of the North Atlantic region, and is expected to lead to generally negative economic and societal knock-on effects, especially in Europe but also globally \cite{ritchie20,wunderling21,orihuela22}. 

A variety of models have been used to study the dynamics of the AMOC, from simplified conceptual box models \cite{STO61, STO68, WEL86}, to complex Earth System Models (ESMs) comprising of millions of variables \cite{LEN09,RAH95}. Sophisticated ESMs offer the most realistic results; however, they also require excessive computational resources. In fact, simpler models, including ocean box models, have been vital for improving our understanding of the fundamental dynamics of the AMOC, and assist in interpreting the results of the more complex models.  

\begin{figure}[t]
    \centering
    \includegraphics{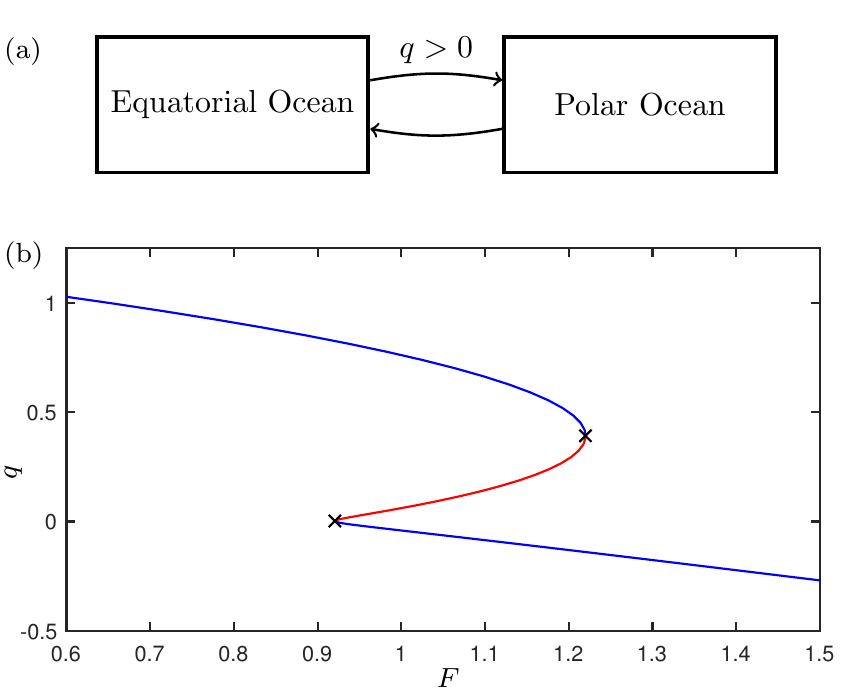}
    \caption{Schematic diagram of ocean circulation (a) and bifurcation diagram (b) for the Stommel two-box model. Circulation direction is shown for $q>0$. Direction is reversed for $q<0$. Parameter $F$ represents a freshwater flux into the polar box. Blue/red curves show branches of stable/unstable equilibria. Black crosses denote fold bifurcations. 
    }
\label{fig:stommel}
\end{figure}

Henry Stommel pioneered the first known ocean box model in 1961 \cite{STO61}. The model consists of two boxes, which could be used to represent equatorial and polar oceanic reservoirs, each with dynamic variables representing salinity and temperature. Those variables define a density gradient between the two boxes and are used to prescribe a dynamic rate of circular water flow $q$ between the boxes. The magnitude $q$ measures the circulation strength, while the sign of $q$ dictates whether the flow between the boxes is clockwise or anti-clockwise; see the diagram in Fig.~\ref{fig:stommel}(a).
Interestingly, Stommel proved the existence of multiple equilibria under certain parameter constraints. In particular, it is possible for states of positive $q$ and negative $q$ to co-exist for the same set of parameters: the observed state depends only on the initial conditions. A bifurcation analysis of the Stommel model (for example, see \cite{OLB12}) reveals two stable branches (one $q>0$ and one $q<0$) when the rate of freshwater forcing into one of the boxes is a bifurcation parameter. These stable branches are connected via fold bifurcations to an unstable branch, such that a region of bistability exists for a range of freshwater forcing rates between the states with $q>0$ and $q<0$. 
This is illustrated in Fig.~\ref{fig:stommel}(b), which shows the bifurcation diagram of a nondimensionalised version of the Stommel two-box model (see \ref{app:stommel} for details). The bifurcation parameter $F$ represents a freshwater flux into the polar box. Blue and red curves are stable and unstable branches equilibria, respectively, which meet at fold bifurcations, denoted by black crosses. The upper and lower stable branches correspond to $q>0$ and $q<0$ equilibria, respectively, such that a region of bistability between the two different circulation patterns exists.

Pierre Welander expanded the box model concept further in 1986 with the aim of studying interhemispheric circulation patterns \cite{WEL86}. His model comprises three boxes: one equatorial box, one polar box in the North, and one polar box in the South. Welander’s model is used to investigate the dynamics of meridional flow across an entire ocean basin, for example, the Atlantic Ocean. This three-box model permits the existence of up to nine equilibria and four stable circulation patterns (to be discussed in detail below).

In more recent years, ocean box models have been further developed and analysed in order to study more complex interactions and phenomena. For example, different mechanisms for millennial-scale oscillations in circulation strengths have been studied in Refs.~\cite{deVerdiere07, keane22}. Rate-induced tipping in a 5-box model calibrated to simulations of a GCM was demonstrated in Ref.~\cite{alkhayuon19}. In Refs.~\cite{rahmstorf01,kuhlbrodt01} the effects of seasonal and stochastic forcing on overturning strength was investigated with a box model calibrated to observational data from the Labrador Sea.

In this study, we consider an adaptation of the Welander three-box model to investigate the effects of two distinct deep-water formation sites in the North Atlantic Ocean. To date, deep-water formation in the North Atlantic Ocean has been represented by only one box, or sometimes two (an upper and lower box) when modelling deep-water formation. In other words, there is a single deep-water formation site, usually located in the Nordic Seas. However, deep-water formation in the Labrador Sea, in addition to in the Nordic Seas, is believed to play an important role in AMOC variability \cite{RAH95,yang16,koenigk21}, albeit not necessarily the most important role as we now know from the OSNAP 
array \cite{lozier19}. 
Therefore, in the present study, we use two boxes to represent two different geographical sites of deep-water formation in the North Atlantic Ocean. We show that an understanding of the resulting dynamics can help to provide insight into what a transition to an AMOC shutdown state may look like.

In the following section we provide a detailed description of the model. In Section~\ref{sec:bif} we study the effects of different freshwater fluxes and different thermal forcings on the two polar boxes by means of bifurcation analysis. We find that the asymmetry between the polar boxes leads to additional bifurcations, such that the tipping of AMOC is not necessarily characterised by a single bifurcation. The results of our analysis are shown to be in agreement with features in simulations from more complex ESMs in Section~\ref{sec:fancy}. These features include a step slowdown preceding the collapse of the AMOC, additional multistability during the tipping event and an intermittent transition as the deep-water formation in the Labrador Sea shuts down. A discussion of the results, and possible implications, concludes the paper in Section \ref{sec:discussion}.

\section{Model description}
\label{sec:model}

\begin{figure}[t]
    \centering
    \includegraphics[width=\columnwidth,trim= 0 0 0 0,clip]{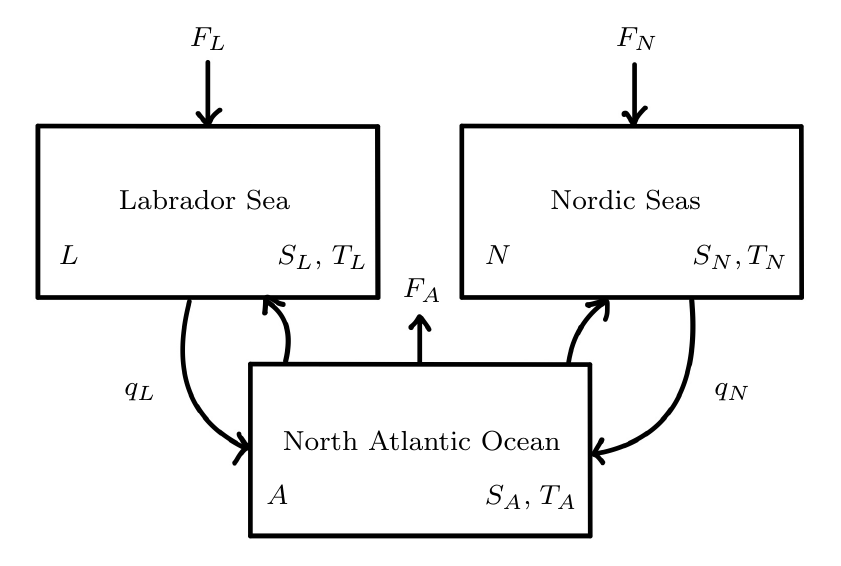}
    \caption{Schematic diagram of the three-box model. Each box represents a region of ocean with salt variables $(S_L,S_N,S_A)$, temperature parameters $(T_L,T_N,T_A)$ and freshwater flux parameters $(F_L,F_N,F_A)$. The arrows between the boxes represent circulation of strengths $q_L$ and $q_N$. The arrow directions assume positive $q_L$ and $q_N$.}
    \label{fig:my_welander}
\end{figure}

%Schematic - Figure~\ref{fig:my_welander}
Figure~\ref{fig:my_welander} shows a schematic diagram of the three-box ocean model we consider in this paper. Box~A corresponds to the mid-latitude North Atlantic Ocean, box~L corresponds to the Labrador Sea, and box~N corresponds to the Nordic Seas (Greenland, Icelandic and Norwegian Seas). Each box has a prescribed constant temperature $(T_A,T_L,T_N)$ and a dynamic salinity variable $(S_A,S_L,S_N)$. These properties determine density gradients between boxes~A and~L, and boxes~A and~N, which induce density-driven circulation of strengths $q_L$ and $q_N$, respectively. The parameters $(F_A,F_L,F_N)$ represent freshwater fluxes. Generally, freshwater leaves box~A by evaporation and enters boxes~L and~N by precipitation.

%Equations
The governing equations of the model are:
\begin{align}
    \label{eq:model_nominal}
    V_L \dot{S_L} & = -F_L S_0 + |q_L| (S_{A} - S_L),  \nonumber\\
    V_N \dot{S_N} & = -F_N S_0 + |q_N| (S_{A} - S_N), \nonumber\\
    V_{Total}S_{0} & = V_L S_L + V_{A} S_{A} +V_N S_N, \\
    q_L & = -k\left[\alpha(T_L-T_{A}) -\beta(S_L -S_{A})\right], \nonumber\\
    q_N & = -k\left[\alpha(T_N-T_{A}) -\beta(S_N -S_{A})\right], \nonumber
\end{align}
with $V_{Total} = V_L + V_N + V_A$ (the sum of all box volumes). The parameter $S_0$ is a reference salinity, parameter $k$ is a hydraulic constant, and $\alpha$ and $\beta$ are temperature and salinity expansion coefficients, respectively. 
The model is two-dimensional because we assume the conservation of salt, such that $S_A=S_{Total}-S_L-S_N$, where $S_{Total}=3S_0$.
Note that for the special case of $F_L=F_N$, $T_L=T_N$ and $V_L=V_N=V_A$, we recover the equations of the original Welander three-box model with symmetric (North/South) polar boxes. 

Here, we use the common modelling assumption that the temperatures of each box is a fixed constant given by a parameter rather than a dependent variable \cite{WEL86,OLB12,alkhayuon19,TIT02}. This approximation is based on the fact that surface ocean temperature anomalies are damped much faster by the atmosphere processes  compared to salinity anomalies, so that the ocean already reaches thermal equilibrium while salinity still adjusts.

%Parameters
Although we study a mathematically very simple box model in order to understand \emph{qualitative} behaviour, and not necessarily quantitative behaviour, we do want the asymmetries between the polar boxes of the model to be sufficiently realistic. 
Therefore, we choose freshwater flux parameter values obtained from the ERA-20C dataset \cite{poli16}. The temperature parameters are calculated from the Mercator Ocean re-analysis data \cite{ferry10}. These parameters, as well as the volume parameters, are determined for the Labrador Sea (64\degree W - 43\degree W, 47\degree N - 60\degree N), the Nordic  Seas (30\degree W - 5\degree E, 66\degree N - 78\degree N) and the mid-latitude North Atlantic between 15\degree N - 45\degree N. We assume an average depth of 2km for the Labrador and Nordic Seas, and 4km for the mid-latitude North Atlantic Ocean. The nominal parameter values are summarised in Table~\ref{table:par}.
As is common practice for such ocean box models, once all other parameters values are chosen, we choose a value for $k$ that gives us sufficiently realistic circulation strengths.

\begin{table}[t]
\footnotesize
\begin{center}
\begin{tabular}{r|l||r|l}
$V_A$ & $6.3515\times 10^{16} \text{ m}^{3}$ & $T_A$    & $5.554$ \degree C   \\
$V_N$ & $3.2036\times 10^{15} \text{ m}^{3}$ & $T_N$    & $0.175$ \degree C  \\
$V_L$ & $4.0070\times 10^{15} \text{ m}^{3}$ & $T_L$    & $3.018$ \degree C  \\
$S_0$ & $35$ psu                              & $F_N$ & $0.0166$ Sv  \\
$\alpha$ & $1.7 \times 10^{-4} \text{ K}^{-1}$       & $F_L$  & $0.0322$ Sv \\
$\beta$  & $0.8 \times 10^{-3}\text{ psu}^{-1}$      & $k$  & $2.1\times 10^4 \text{ Sv}$ 
\end{tabular}
\end{center}
\vspace*{-5mm}
\normalsize
\caption{Nominal parameter values.}
\label{table:par}
\end{table}

\begin{figure}
    \centering
    \includegraphics[width=\columnwidth,trim= 0 0 0 0,clip]{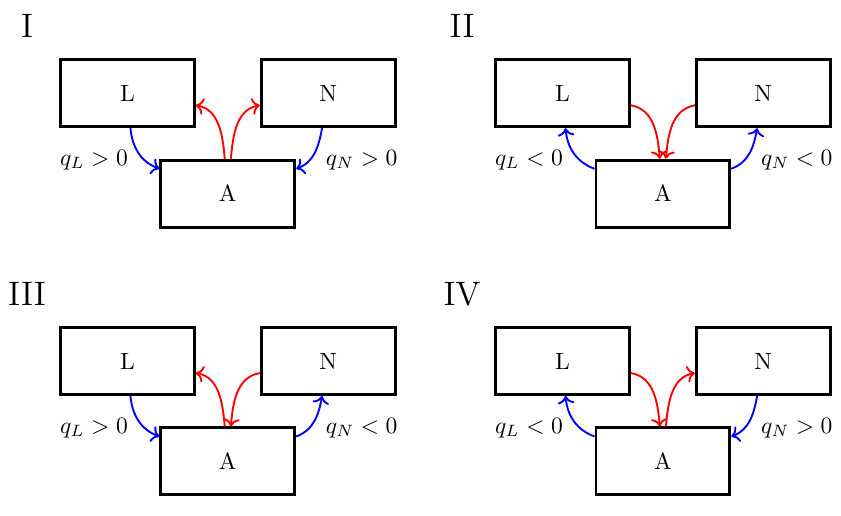}
    \caption{Schematic of circulation patterns between the three boxes for different signs of $q_L$ and $q_N$. Red/blue arrows represent warm/cold surface/deep water transport.
    }
    \label{fig:circulation}
\end{figure}

An analysis of the Welander three-box model reveals the existence of up to nine equilibria \cite{WEL86,OLB12}. The same is true for model~(\ref{eq:model_nominal}), as described in the analysis in the following section. Out of these nine equilibria, four are stable and represent qualitatively different circulation patterns, which are shown in Fig.~\ref{fig:circulation}. Patterns~I--IV illustrate flow between box~A and~L, and box~A and~N, depending on the signs of $q_L$ and $q_N$. Red and blue arrows represent the flow of warm surface water and cold deep water, respectively.
In pattern~I both $q_L$ and $q_N$ are positive, such that the warm surface water travels northward to the polar boxes and deep-water formation occurs in both polar boxes (representative of modern circulation in the Atlantic Ocean). In contrast, pattern~II has negative $q_L$ and $q_N$, such that warm surface water travels southward. In the model, this equilibrium implies that deep-water formation occurs in box~A (the mid-latitude North Atlantic Ocean), which is not the present state. We could adapt the model so that $q_L$ and $q_N$ are either positive or zero. Instead, we simply consider this circulation pattern to represent deep-water formation shutting down in both polar regions. Finally, patterns~III and~IV each represent patterns where only one deep-water formation site is active.

We are interested in understanding the effects of asymmetry between boxes~L and~N.
In Ref.~\cite[p.524]{OLB12} is was shown how introducing an asymmetric freshwater flux to the Welander three-box model breaks the symmetry of the solutions. Here, we investigate an asymmetric freshwater flux in the context of differentiating deep-water formation in the Labrador Sea and Nordic Seas. We also allow asymmetric thermal forcing between the two regions. 
Hence, we are not interested in the fact that symmetry is broken but in how this affects the observed behaviour.
%Homotopy
To this end, we introduce homotopy transitions from a symmetric model to an asymmetric model via parameters $\eta$ and $\mu$ to write system~(\ref{eq:model_nominal}) in the form:
%Equations
\begin{align}
    \label{eq:model}
    V_L \dot{S_L} & = -(F+F_N+\eta(F_L-F_N) )S_0, \nonumber\\ &+ |q_L| (S_{A} - S_L),  \nonumber\\
    V_N \dot{S_N} & = -(F+F_N) S_0 + |q_N| (S_{A} - S_N), \nonumber\\
    V_{Total}S_{0} & = V_L S_L + V_{A} S_{A} +V_N S_N, \\
    q_L & = -k[\alpha(T_N+\mu(T_L-T_N)-T_{A}) \nonumber\\&-\beta(S_L -S_{A})], \nonumber\\
    q_N & = -k\left[\alpha(T_N-T_{A}) -\beta(S_N -S_{A})\right]. \nonumber
\end{align}
When $\eta=0$ and $\mu=0$ both boxes~L and~N have freshwater flux $F_N$ and thermal forcing $T_N$, so that the roots of the right-hand sides of the $S_L$ and $S_N$ equations are identical. When $\eta=1$ and $\mu=1$ box~L takes on its nominal values $F_L$ and $T_L$, respectively.

Note the introduction of parameter $F$, which represents additional freshwater flux from box~A into boxes~L and~N. Such a parameter is typically introduced to ocean models (across the complexity spectrum) in order to conduct so-called freshwater hosing experiments, where the increase in freshwater flux reflects global warming effects. It will serve as our bifurcation parameter.

\section{Bifurcation analysis} 
\label{sec:bif}

We first analyse the symmetric case of model~(\ref{eq:model}) with $\eta=\mu=0$, then consider the effects of increasing $\eta$ and $\mu$ to $1$, so that each box has its nominal parameter values (given in Table~\ref{table:par}).
Throughout this section we show bifurcation diagrams with $F$ as the bifurcation parameter, displayed in terms of the variables $S_L$ and $S_N$, as well as the total circulation strength, given by $Q=q_N+q_L$. In our model $Q$ is the observable, and it is comparable to the AMOC strength that is calculated as an observable in more sophisticated ESMs.

\subsection{Symmetric case}
\label{subsec:bif_sym}

\begin{figure}[t]
    \centering
    \includegraphics{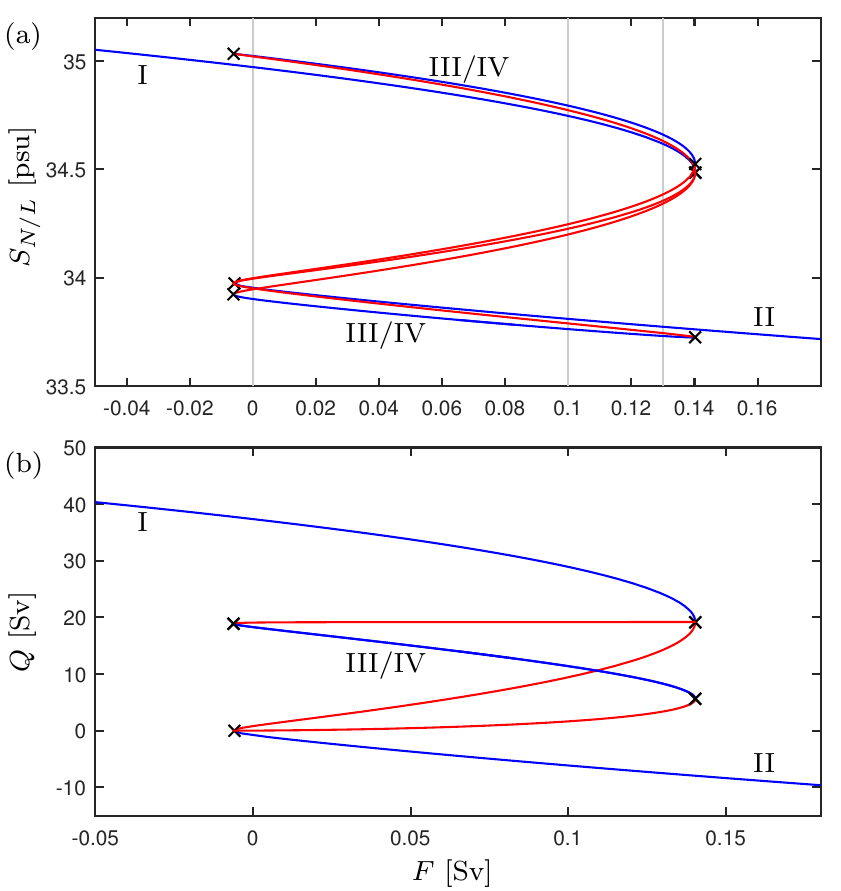}
    \caption{Bifurcation diagram for symmetric case
    $(\eta,\mu)=(0,0)$ in terms of freshwater flux $F$ and $S_N=S_L$ in panel~(a) and total circulation strength $Q$ in panel~(b). Note that, due to symmetry, the bifurcation diagrams of $S_N$ and $S_L$ are identical. Blue/red curves show branches of stable/unstable equilibria. Black crosses denote fold bifurcations. Roman numerals refer to the circulation patterns in Fig.~\ref{fig:circulation}. Grey lines in panel~(a) correspond to values of $F$ used in Fig.~\ref{fig:phase_sym}. 
    }
\label{fig:bif_sym}
\end{figure}

Figure~\ref{fig:bif_sym} shows the bifurcation diagram for the symmetric case, calculated with numerical continuation software \cite{ENG01,SIE14}.
It shows several co-existing branches of stable and unstable equilibria, in blue and red, respectively. Fold bifurcations are marked by black crosses. In panel~(a) we observe the effects of the symmetry between variables $S_L$ and $S_N$. For some branches, $S_L$ and $S_N$ are equal and for the others they are symmetrically related; more specifically, if the equilibrium solution $(S_L,S_N)=(x_1,x_2)$ exists, then so does $(x_2,x_1)$. We also see that multiple fold bifurcations occur simultaneously as $F$ passes near $-0.01$ and $0.14$. The Roman numerals label the stable branches according to the circulation patterns shown in Fig.~\ref{fig:circulation}. Along branch I both $S_N$ and $S_L$ are equal and are sufficiently large such that both deep-water formation sites are active. Along branch II they are both equal and sufficiently small such that both sites are inactive. The other stable branches correspond to patterns III--IV, where $S_N$ and $S_L$ are not equal, and in each case only one deep-water formation sites is active.

\begin{figure}
    \centering
    \includegraphics[width=0.9\columnwidth]{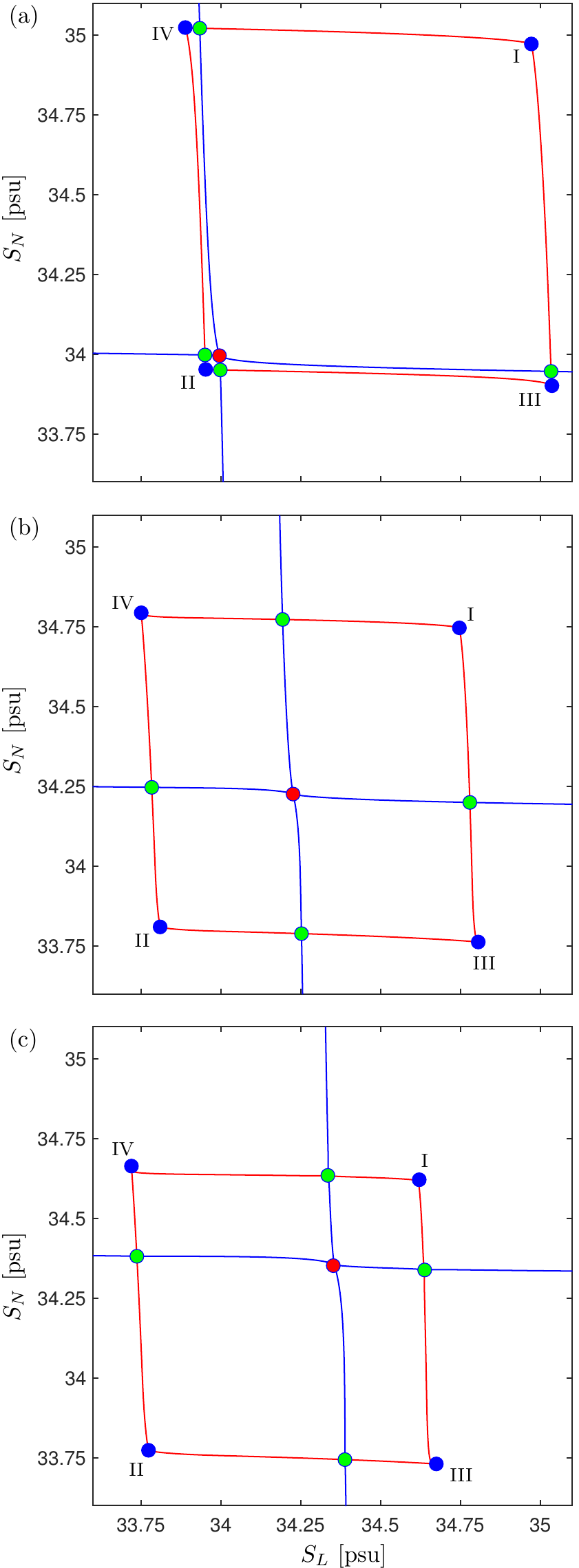}
    \caption{Phase plane diagrams for the symmetric case $(\eta,\mu)=(0,0)$ with $F=0$ (a), $F=0.1$ (b) and $F=0.13$ (c). Blue/green/red dots represent stable/saddle/unstable equilibria. Blue/red curves show numerically computed stable/unstable manifolds of the saddles. Roman numerals refer to the circulation patterns in Fig.~\ref{fig:circulation}.
    }
    \label{fig:phase_sym}
\end{figure}

Figure~\ref{fig:bif_sym}(b) shows the effect of varying $F$ on the circulation strength $Q$, to illustrate what to expect from a freshwater hosing experiment. For low values of $F$, we begin on the upper branch with large $Q$ [pattern I]. As $F$ increases, the circulation strength becomes weaker until a fold bifurcation is met and the system transitions to the lower branch with both deep-water formation sites inactive [pattern II]. When $F$ is decreased, the system eventually undergoes another fold bifurcation and transitions back to the upper branch. Meanwhile, there are two other overlapping branches of stable equilibria that would not be encountered during a freshwater hosing experiment, unless the starting values of $F$ and initial conditions are chosen very specifically. These two branches each represent the states where only one deep-water formation site is active [patterns III and IV].

Figure~\ref{fig:phase_sym} displays phase plane diagrams for the values of $F$ indicated by grey lines in Fig.~\ref{fig:bif_sym}(a). Blue, green and red circles represent stable, saddle and unstable equilibria, respectively. Each blue circle corresponds to a particular circulation pattern as labelled by the Roman numerals. The blue and red curves are approximations of the stable and unstable manifolds of the saddles; they are found by numerical integration of the model forwards and backwards in time from near the saddles and along the respective eigendirections. A striking feature of Fig.~\ref{fig:phase_sym} is the symmetry across the diagonal of the plane, due to the exchanging symmetry between $S_L$ and $S_N$ for $\mu=\eta=0$. The panels also provide a clear impression of the fold bifurcations that take place. Panel~(a) shows the phase plane very shortly after three simultaneous fold bifurcations have taken place near $F=-0.01$. Two fold bifurcations are the meeting of the stable nodes III and IV with their respective saddles. The third is a degenerate fold bifurcation on the line of symmetry where the stable node II, an unstable node and two saddles coalesce. 

In Fig.~\ref{fig:phase_sym}(b) all equilibria have moved away from each other. In panel~(c) they have moved closer together again in different groupings en route to the three fold bifurcations that occur near $F=0.14$. Here all equilibria apart from stable node II will disappear.

\subsection{Asymmetric case}
\label{subsec:bif_asym}

\begin{figure}[t]
    \centering
    \includegraphics{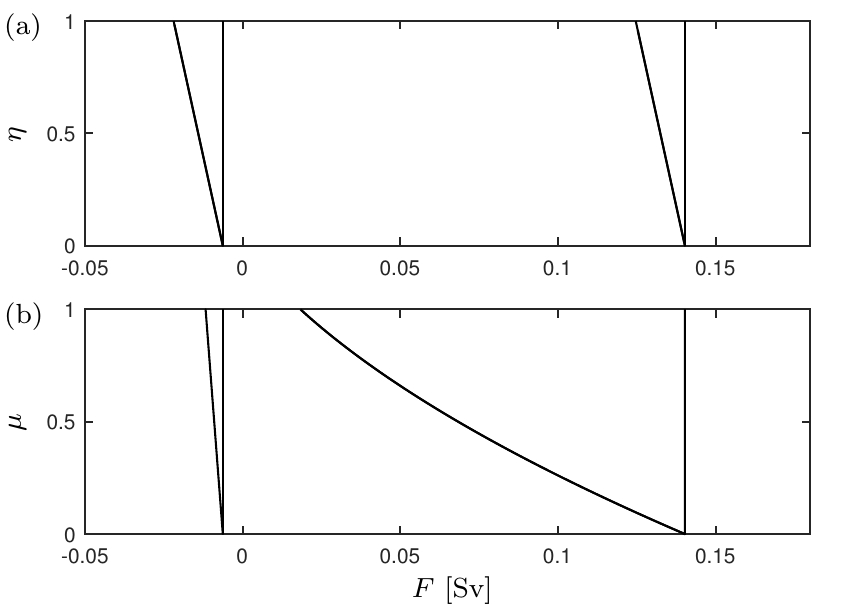}
    \caption{Curves of fold bifurcations of the box model in the (a) $(F,\eta)$-plane and (b) $(F,\mu)$-plane.
    }
    \label{fig:folds}
\end{figure}

Now we consider the effect of introducing asymmetry between the boxes $L$ and $N$ in terms of freshwater flux and thermal forcing. In other words, we increase $\eta$ and/or $\mu$ in model~(\ref{eq:model}) from zero. Figure~\ref{fig:folds} shows curves of all the fold bifurcations in the $(F,\eta)$-plane (a) and the $(F,\mu)$-plane (b). 
%Some of the curves overlap. 
In each panel the curves begin near $F=-0.01$ and $0.14$, as observed in the previous section. Once $\eta$ equals $1$ in panel~(a), box~$L$ attains its nominal value for $F_L$ given in Table~\ref{table:par}. Similarly, in panel~(b) once $\mu$ equals $1$, box~$N$ attains its nominal value for $T_L$. We see that in both cases some fold bifurcations are affected by the change, while some are not and appear as purely vertical lines. We also observe that the freshwater asymmetry appears to have a stronger effect on fold bifurcations occurring for negative $F$, while thermal asymmetry has a significantly stronger effect on fold bifurcations occurring for positive $F$. 

\begin{figure}[t]
    \centering
    \includegraphics{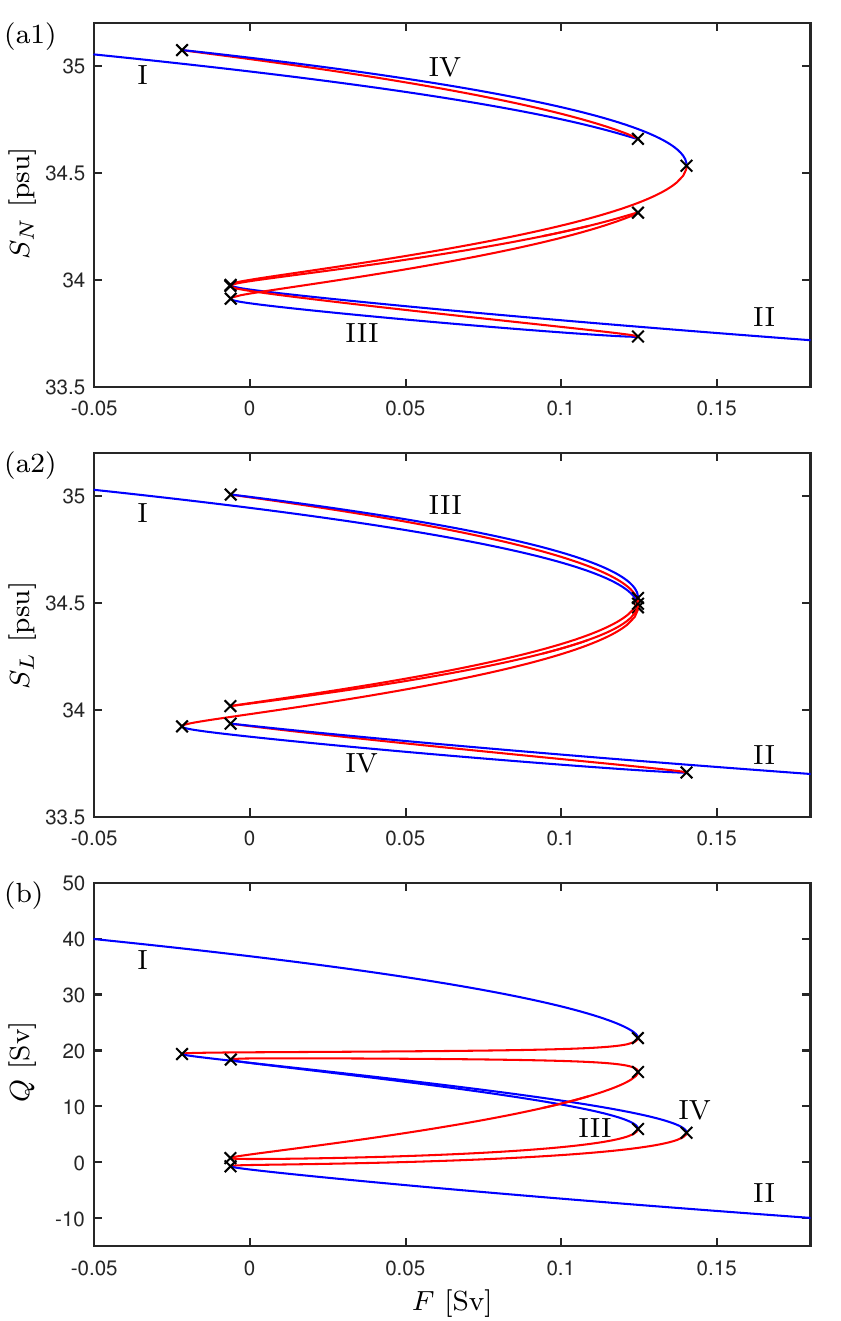}
    \caption{Bifurcation diagram for the asymmetric case $(\eta,\mu)=(1,0)$ in terms of freshwater flux $F$ and $S_N$ in panel~(a), $S_L$ in panel~(b) and $Q$ in panel~(c). Blue/red curves show branches of stable/unstable equilibria. Black crosses denote fold bifurcations. Roman numerals refer to the circulation patterns in Fig.~\ref{fig:circulation}.
    }
    \label{fig:bif_asymF_symT}
\end{figure}

In Fig.~\ref{fig:bif_asymF_symT} we show the bifurcation diagrams for the case of asymmetric freshwater flux and symmetric thermal forcing (i.e. $(\eta,\mu)=(1,0)$). As expected, the bifurcation diagrams in terms of $S_L$ and $S_N$ in panels~(a1) and~(a2), respectively, are no longer symmetrically related. Due to this asymmetry, fold bifurcation occur at four distinct values of $F$, as already indicated at the top of panel~(a) of Fig.~\ref{fig:folds}. 
This has implications as we consider the observable $Q$ in dependence on $F$. As seen in Fig.~\ref{fig:bif_asymF_symT}(b), when the strong circulation pattern I (with large $Q$) weakens during an increase in $F$ and undergoes a critical transition at a fold bifurcation near $F=0.125$, it does not immediately transition to the state with both deep-water formation sites inactive. Instead, as $F$ increases the system encounters the intermediate state with deep-water formation only active in box~$N$ [pattern IV] before finally transitioning to branch II near $F=0.14$. Similarly, when the lower branch terminates near $F=-0.01$ as $F$ decreases in Fig.~\ref{fig:bif_asymF_symT}(b), the same intermediate state is encountered before transitioning to the upper branch.

\begin{figure}[t]
    \centering
    \includegraphics{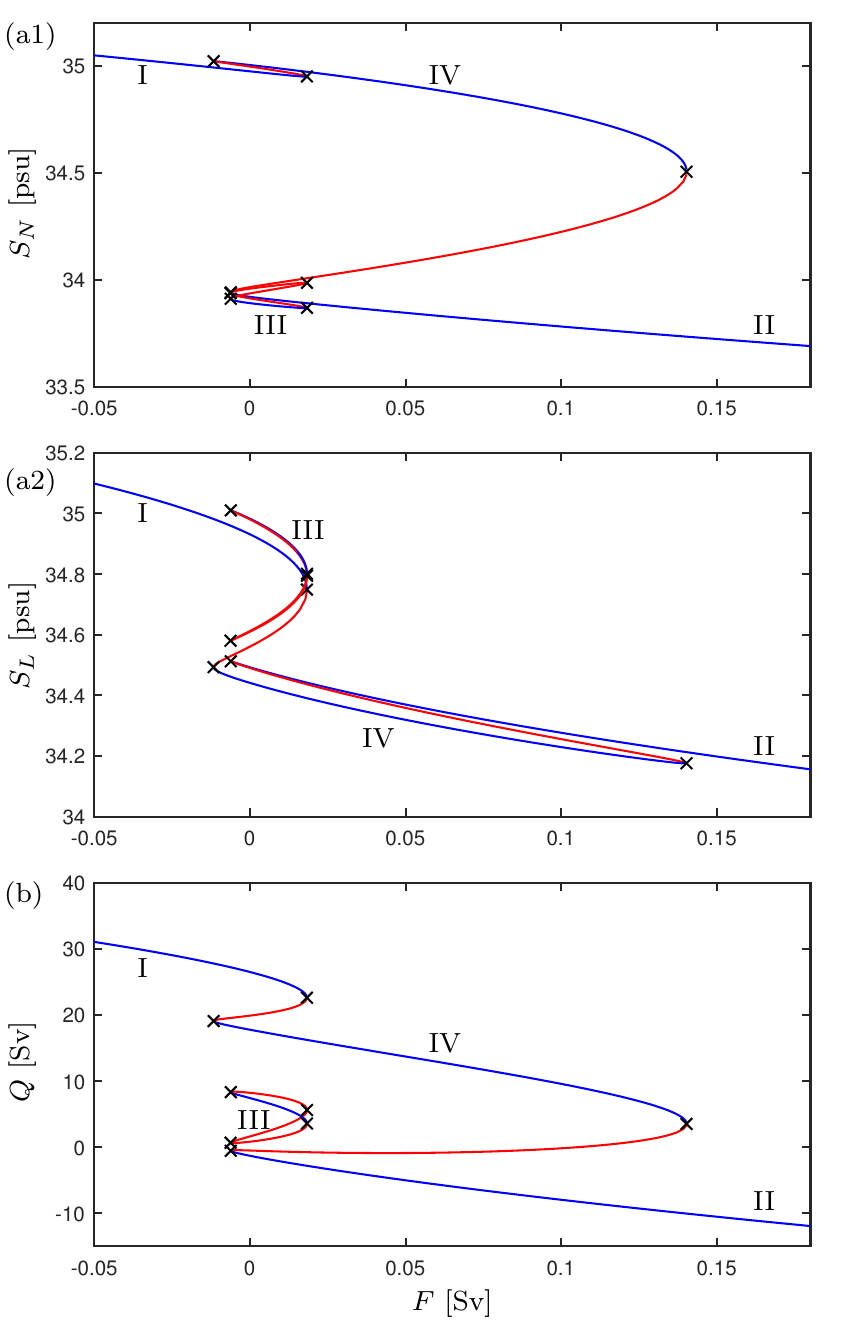}
    \caption{Bifurcation diagram for the asymmetric case $(\eta,\mu)=(0,1)$ in terms of freshwater flux $F$ and $S_N$ in panel~(a), $S_L$ in panel~(b) and $Q$ in panel~(c). Blue/red curves show branches of stable/unstable equilibria. Black crosses denote fold bifurcations. Roman numerals refer to the circulation patterns in Fig.~\ref{fig:circulation}.}
    \label{fig:bif_symF_asymT}
\end{figure}

Figure~\ref{fig:bif_symF_asymT} shows the complementary case of asymmetric thermal forcing (i.e. $(\eta,\mu)=(0,1)$). Similar branches are observed in this case, however the effect of shifting the fold bifurcation locations in terms of $F$ is stronger. One important implication is that branch I loses stability at a lower value of $F$. Therefore, as seen in Fig.~\ref{fig:bif_symF_asymT}(b), if the system transitions from branch I as $F$ increases, the range of values of $F$ that the system stays in the intermediate state with one active deep-water formation site [pattern IV] is much larger. Another effect that is clearer to see here is that the branch of equilibria to which the other intermediate state [pattern III] belongs exists as an isola and cannot be observed without carefully chosen initial conditions.

\begin{figure*}
    \centering
    \includegraphics{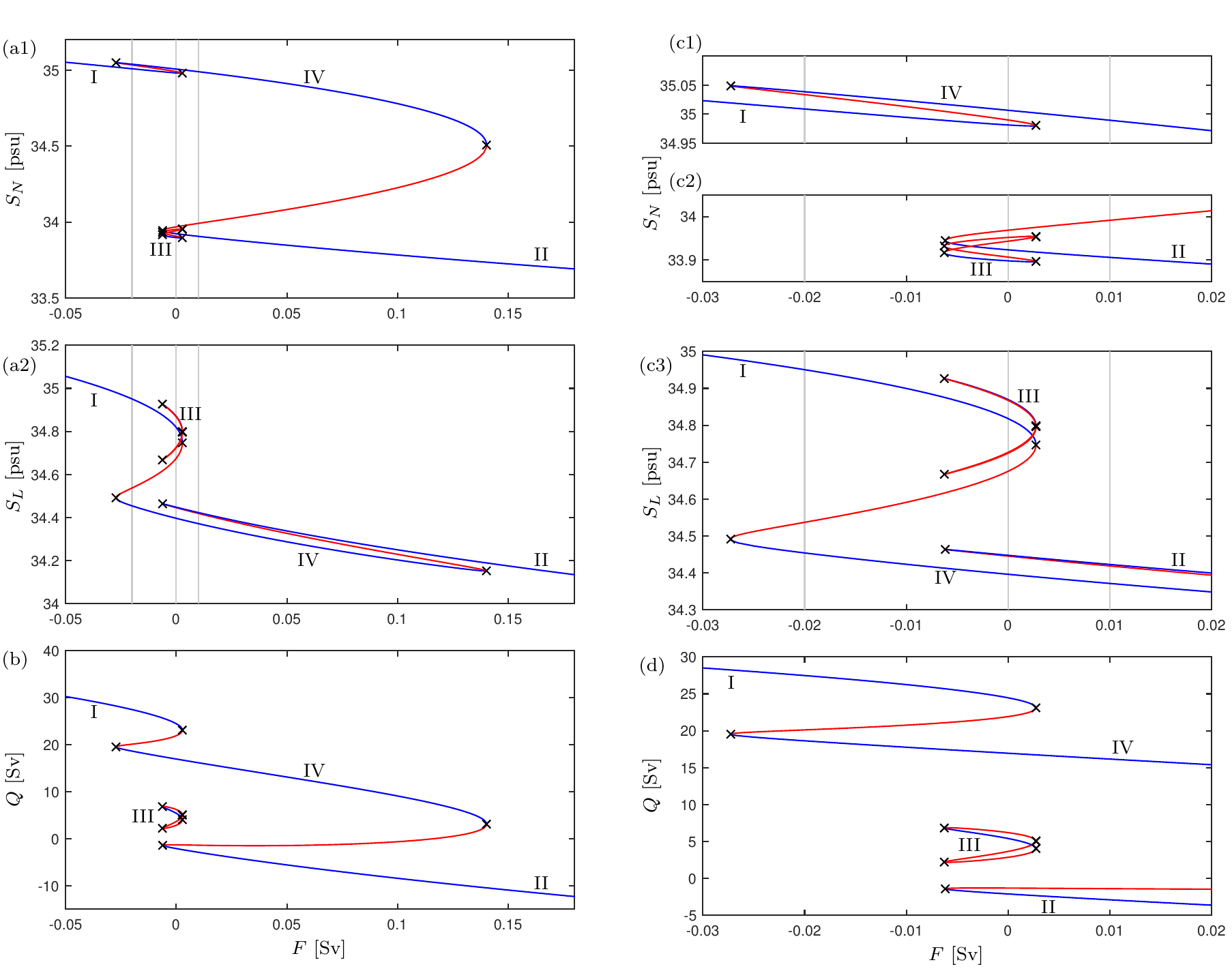}
    \caption{Bifurcation diagram for the asymmetric case $(\eta,\mu)=(1,1)$ in terms of freshwater flux $F$ and $S_N$ in panel~(a), $S_L$ in panel~(b) and $Q$ in panel~(c). Blue/red curves show branches of stable/unstable equilibria. Black crosses denote fold bifurcations. Roman numerals refer to the circulation patterns in Fig.~\ref{fig:circulation}.
    Panels~(c)--(d) provide close-up plots of the bifurcations in panels~(a)--(b).
    }
    \label{fig:bif_asymF_asymT}
\end{figure*}

Finally, we consider the case of both asymmetries in model~(\ref{eq:model}) with parameters at their nominal values (i.e. $(\eta,\mu)=(1,1)$). Figure~\ref{fig:bif_asymF_asymT} shows bifurcation diagrams with additional close-up plots on the right-hand side. Panels~(a)--(b) appear to be very similar to Fig.~\ref{fig:bif_symF_asymT}, reflecting that the asymmetry in thermal forcing makes the larger contribution to the effects on the bifurcation diagrams compared to the asymmetry in freshwater flux. 

\begin{figure}[t]
    \centering
    \includegraphics{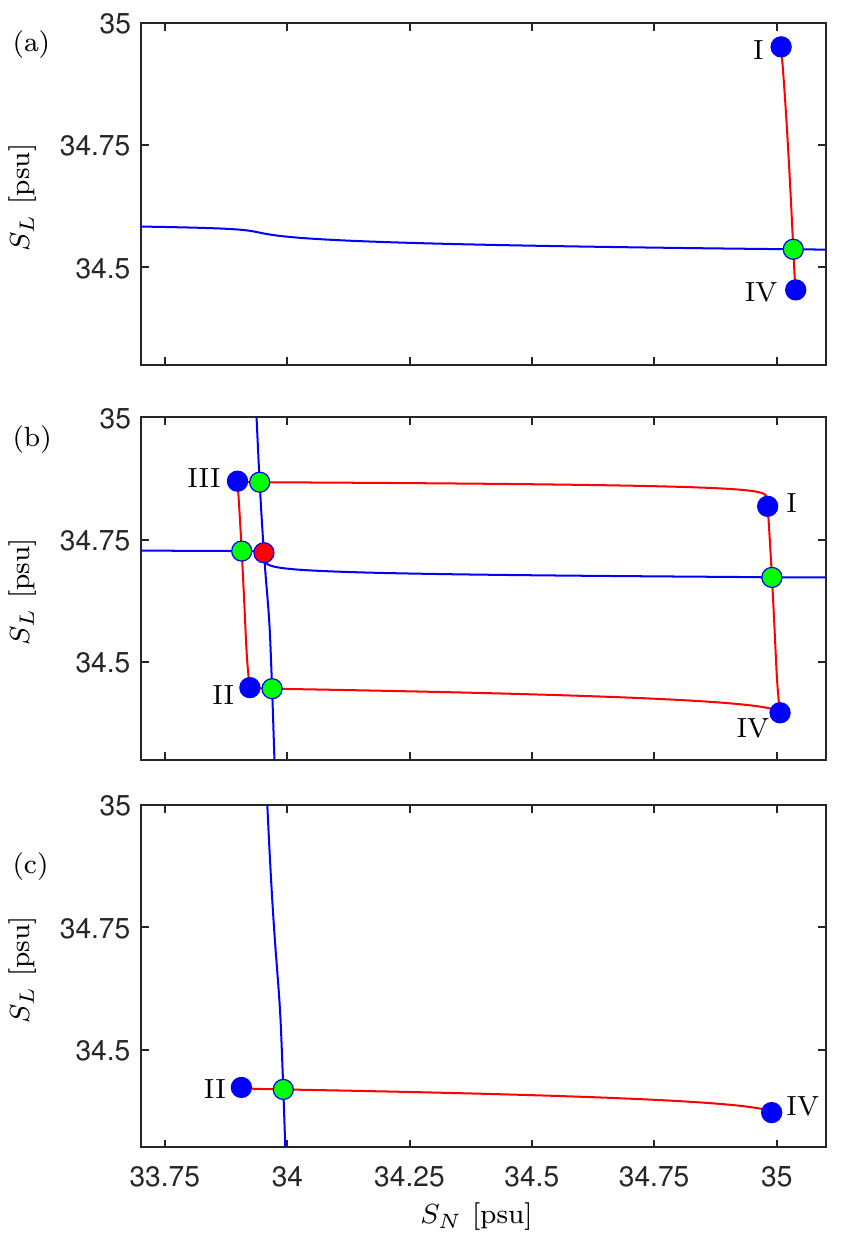}
    \caption{Phase plane diagrams for symmetric case $(\eta,\mu)=(1,1)$ for $F=-0.02$ (a), $F=0$ (b) and $F=0.01$ (c). Blue/green/red dots represent stable/saddle/unstable equilibria. Blue/red curves show numerically computed stable/unstable manifolds of the saddles. Roman numerals refer to the circulation patterns in Fig.~\ref{fig:circulation}.
    }
    \label{fig:phase_asymF_asymT}
\end{figure}

Figure~\ref{fig:phase_asymF_asymT} provides phase plane diagrams for the values of $F$ indicated by the grey lines in Fig.~\ref{fig:bif_asymF_asymT}. Figure~\ref{fig:phase_asymF_asymT}(a) shows the equilibria that exist for $F=-0.02$. The stable node I belongs to branch I in Fig.~\ref{fig:bif_asymF_asymT}(d). The stable node IV, which represents deep-water formation in box~$N$ only, and the saddle are the equilibria born from the fold bifurcation near $F=-0.028$. In Fig.~\ref{fig:phase_asymF_asymT}(b) we see that the saddle on the right side of the plane has moved up to the upper-right towards stable node I. Furthermore, on the left side of the plane three new pairs of equilibria exist. They have all been created simultaneously at fold bifurcations occurring near $F=-0.007$ (compare with Fig.~\ref{fig:bif_asymF_asymT}(d)). Finally, in Fig.~\ref{fig:phase_asymF_asymT}(c) only three equilibria remain in the lower half of the plane. The other six equilibria undergo pairwise annihilation at three simultaneously occurring fold bifurcations near $F=0.003$. Upon further increase of $F$, the saddle eventually disappears at a fold bifurcation with the stable node IV. This will leave behind the stable node II, which belongs to the stable branch II in Fig.~\ref{fig:bif_asymF_asymT}(b).

\section{Comparison with higher complexity models}
\label{sec:fancy}

The behaviour of the observable $Q$ in model~(\ref{eq:model}) with the nominal values $(\eta,\mu)=(1,1)$ is now compared with the behaviour of the AMOC strength observed in considerably more sophisticated ESMs during freshwater hosing experiments. 
In this section we introduce additive white noise terms to the differential equations so that the model becomes
\begin{align}
    \label{eq:model_noise}
    d{S_L} & = \frac{1}{V_L}\{-(F + F_L) S_0 + |q_L| (S_{A} - S_L)\}dt \nonumber\\
    & + \epsilon dW,  \nonumber\\
    d{S_N} & = \frac{1}{V_N}\{-(F + F_N) S_0 + |q_N| (S_{A} - S_N)\}dt \nonumber\\
    & + \epsilon dW, \nonumber\\
    V_{Total}S_{0} & = V_L S_L + V_{A} S_{A} +V_N S_N, \\
    q_L & = -k\left[\alpha(T_L-T_{A}) -\beta(S_L -S_{A})\right], \nonumber\\
    q_N & = -k\left[\alpha(T_N-T_{A}) -\beta(S_N -S_{A})\right]. \nonumber
\end{align}
The noise is included to represent high-frequency variability present in the more sophisticated ESMs discussed here and, hence, allow a closer comparison. Throughout this section, the noise strength $\epsilon$ is chosen to be sufficiently small such that the behaviour still reflects the underlying bifurcation structure according to the deterministic system.

\begin{figure}[t]
    \centering
    \includegraphics{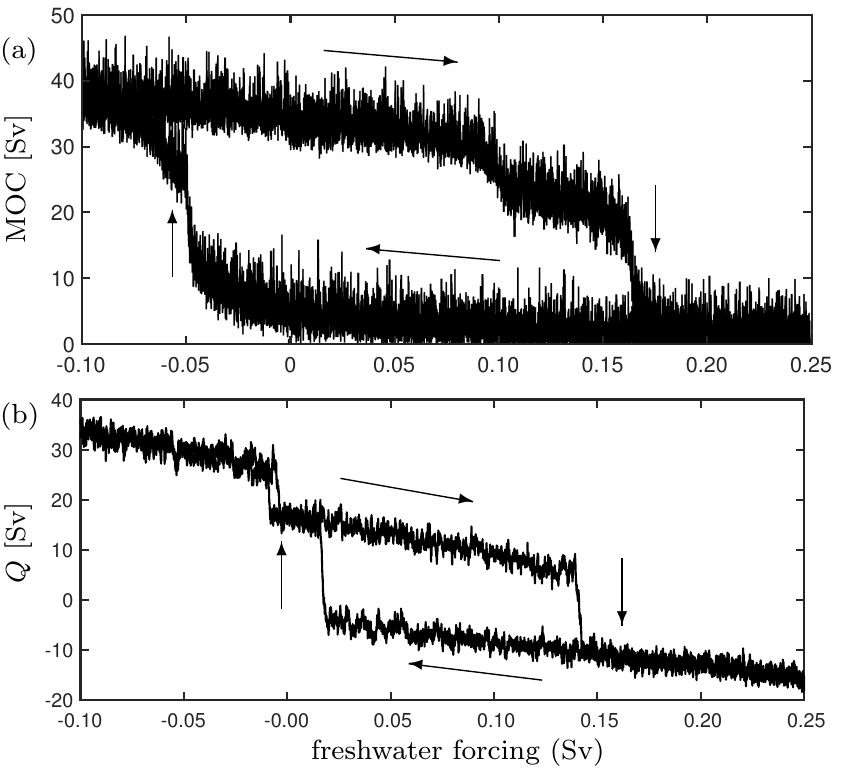}
    \caption{Hysteresis loops: of the GENIE--2 model in panel~(a) as reproduced under Creative Common license (CC BY 4.0) from Ref.~\cite{LEN09}; and of (\ref{eq:model_noise}) in panel~(b) for a drift of $dF/dt = 7 \times 10^{-5}$ Sv/yr and noise strength $\epsilon = 5 \times 10^{-6}\text{~psu}/\sqrt{\text{s}}$.
    }
    \label{fig:genie}
\end{figure}

To begin with, we consider the hysteresis loop created during freshwater hosing experiments by first increasing the rate of freshwater forcing until the strong circulation state is lost, and then decreasing the rate of freshwater forcing until the strong circulation state is recovered.
An example of such a hysteresis loop is provided in Fig.~\ref{fig:genie}(a), reproduced from Ref.~\cite{LEN09}. It shows the maximum overturning circulation (MOC) as freshwater forcing is varied according to the arrows in the intermediate complexity ESM GENIE-2. Interestingly, the authors observed a ``step slowdown'' as a prominent feature in the hysteresis loop, where there is a clear drop in MOC as freshwater forcing is being increased between about $0.10$ and $0.17$ Sv, before the lower branch of the hysteresis loop is reached. 

Figure~\ref{fig:genie}(b) shows a freshwater hosing experiment conducted with model~(\ref{eq:model_noise}), obtained by integration with the Euler-Maruyama method \cite{addrefhere}. The parameter $F$ is slowly varied, according to the arrows, at a rate of $7 \times 10^{-5}$ Sv per year, which is comparable to the rate used in the GENIE-2 experiment. We are not concerned with the fact that the step slowdown occurs in panel~(b) at a smaller level of freshwater forcing, since we are not attempting to reproduce the dynamics of GENIE-2 in a quantitative way. Rather, our knowledge of the three-box model presented in the previous section provides a possible mechanism for the step slowdown observed in panel~(a). Namely, before the shutdown of AMOC near $0.17$ Sv, there is a partial shutdown as the deep-water formation in the Labrador Sea box~L becomes inactive. In model~(\ref{eq:model}) this is due to a fold bifurcation.

There is a slight discrepancy between the two hysteresis loops in  Fig.~\ref{fig:genie}(a) and~(b) in how the system returns to the upper branch. In panel~(b) the intermediate state is briefly visited after the lower branch loses stability, and this does not appear to be the case in panel~(a). This is most likely because the lower branch in panel~(b) and its stability is calculated for the nominal (e.g. surface temperature) parameter values in Table~\ref{table:par}. These parameter values, however, are representative of the modern climate (i.e. the upper branch) and would, realistically, have different values on the lower branch.  

\begin{figure}[t]
    \centering
    \includegraphics{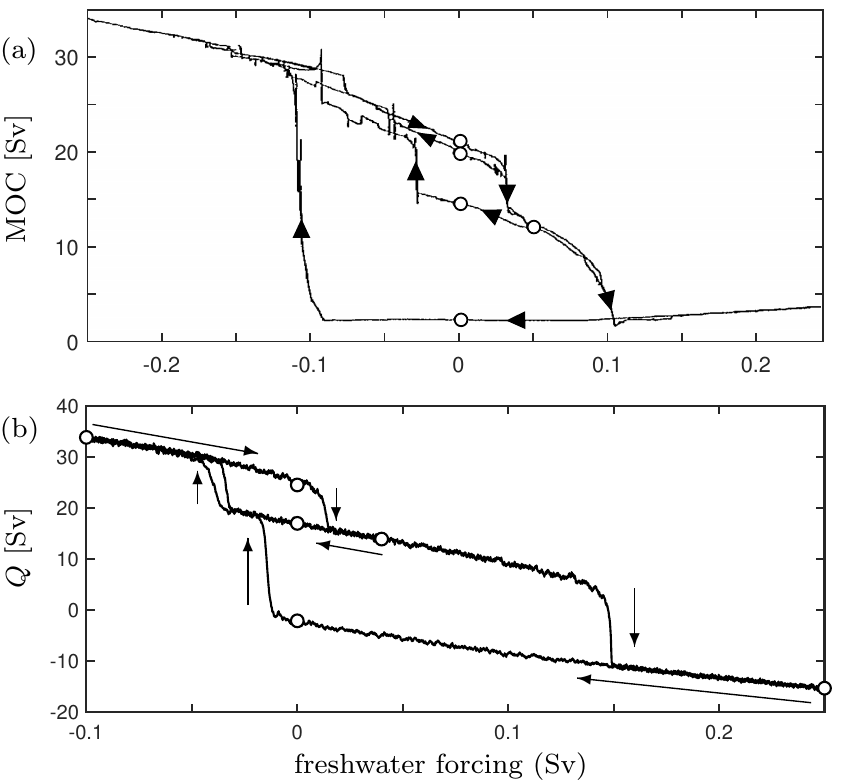}
    \caption{Multistability of tipping event: in a global ocean GCM in panel~(a), reproduced from Ref.~\cite{RAH95} with permission from Springer Nature; and in (\ref{eq:model_noise}) for $dF/dt = 7 \times 10^{-5}$ Sv/yr and noise strength $\epsilon = 1 \times 10^{-6}\text{~psu}/\sqrt{\text{s}}$ in panel~(b).}
    \label{fig:rahmstorf}
\end{figure}

The step slowdown discussed in Fig.~\ref{fig:genie} is in fact a re-occurring feature of such freshwater hosing experiments with sophisticated models; for example, see \cite{ganopolski01,rahmstorf05}. 
Another freshwater hosing experiment that revealed an intermediate circulation state is presented in Ref.~\cite{RAH95}. Here, an ocean GCM coupled with a simplified atmosphere model is used to produce the bifurcation diagram shown in Fig.~\ref{fig:rahmstorf}(a). It shows the hysteresis response of the MOC under varying freshwater forcing, where circles denote states that were given a longer run-time to confirm that they are equilibria. We see that a step slowdown occurs near $0.04$ Sv in panel~(a), which is explicitly shown to coincide with a shutdown of deep-water formation in the Labrador Sea in Ref.~\cite{RAH95}. Furthermore, once the ocean is in the intermediate state near $0.05$ Sv, the freshwater forcing is decreased to reveal the existence of another smaller hysteresis loop. Therefore, there exist at least three stable equilibria at zero freshwater forcing. 
These equilibria represent circulation states with deep-water formation in both the Nordic Seas and the Labrador Sea (on the upper branch $Q\!\approx\!20$ Sv), in the Nordic Seas (on the middle branch $Q\!\approx\!15$ Sv) and no deep-water formation in the northern Atlantic Ocean (on the bottom branch).
%Note that the question of distinguishing the two upper circles in panel~(a) (with MOC values of about 20 and 21 Sv) is discussed in the following section. 

In Fig.~\ref{fig:rahmstorf}(b) the simple box model~(\ref{eq:model_noise}) is shown to reproduce the qualitative behaviour observed in panel~(a). From the equilibria, represented by circles, the freshwater forcing is slowly increased or decreased to reveal branches. As expected, these branches coincide with the stable branches of equilibria shown in Fig.~\ref{fig:bif_asymF_asymT}. From our earlier analysis we also know that the transitions between branches observed in Fig.~\ref{fig:rahmstorf}(b) occur as the stable equilibria disappear at fold bifurcations. This results in two hysteresis loops, in agreement with panel~(a).  Moreover, in panel~(b) three equilibria co-exist at zero freshwater forcing, which represent states with deep-water formation in two, one and zero boxes, analogue to the convection patterns behind the co-existing states in panel~(a).

\begin{figure}[t]
    \centering
    \includegraphics{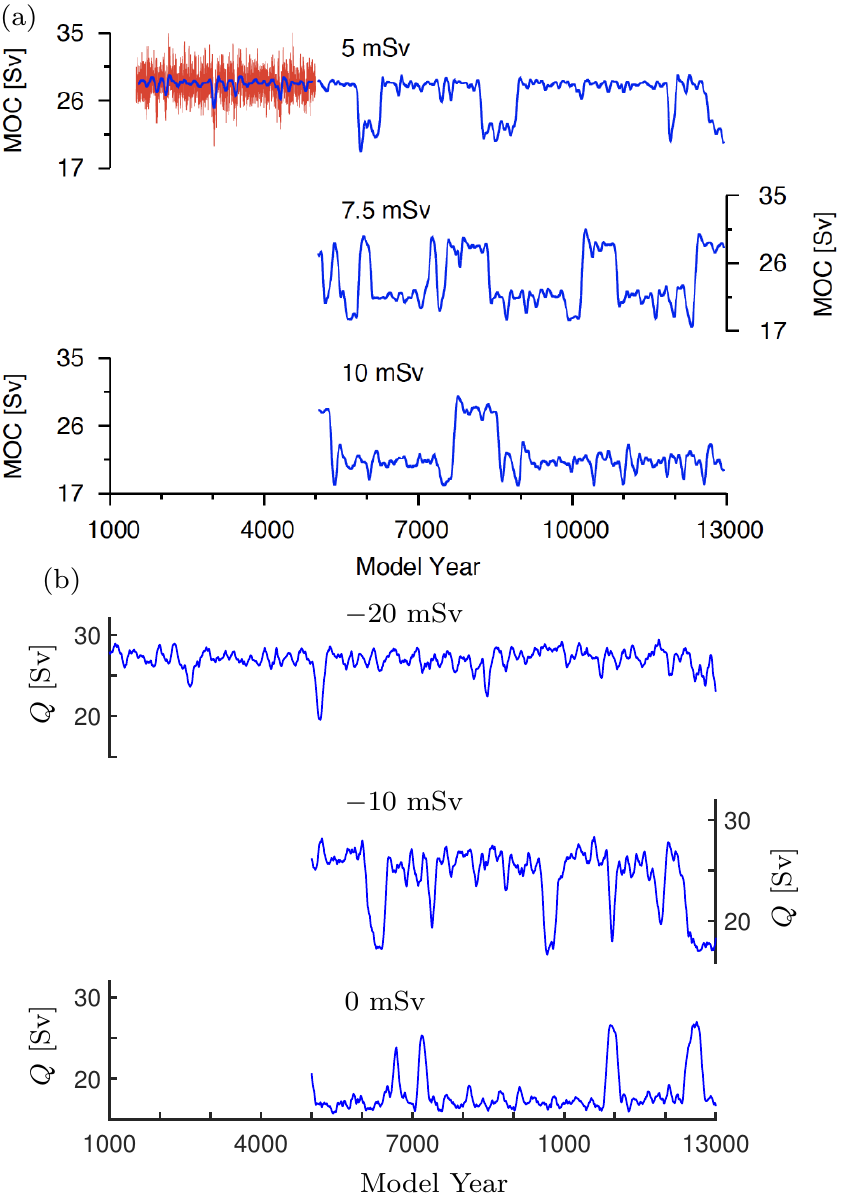}
    \caption{Intermittent transition between stable states: in the ECBilt-CLIO model with freshwater perturbations of 5, 7.5 and 10 mSv at 5000 years in panel~(a), reproduced from \cite{SCH07} under Creative Common license (CC BY-NC-SA 2.5); and in (\ref{eq:model_noise}) for $F=-0.02$, $F=-0.01$ and $F=0$ Sv, and noise strength $\epsilon = 5 \times 10^{-6}\text{~psu}/\sqrt{\text{s}}$ in panel~(b).}
    \label{fig:SCH07}
\end{figure}

In Ref.~\cite{SCH07} the intermediate complexity ECBilt-CLIO model was used to conduct a more precise analysis of the shutdown of deep-water formation in the Labrador Sea due to increasing freshwater forcing.
Figure~\ref{fig:SCH07}(a) shows three experiments that were conducted with different perturbations in freshwater forcing of $5$, $7.5$ and $10$ mSv. The unsmoothed model output (in red) is smoothed by using a $101$-year Hanning filter (in blue). At $5$ mSv the ocean spends most of its time in a state of strong circulation (representative of the upper branches in previous figures), with occasional excursions to a state without deep-water formation in the Labrador Sea (representative of the middle branches in previous figures). The result of increasing freshwater forcing is an intermittent transition with the ocean spending increasingly more time in a state without deep-water formation in the Labrador Sea.

In Fig.~\ref{fig:SCH07}(b) this behaviour is reproduced by the box model~(\ref{eq:model_noise}) with noise. With small increases in $F$ around the transition from the upper branch to the middle branch, we observe an intermittent transition. This provides us with a possible interpretation of the dynamics observed in panel~(a). In the upper half of Fig.~\ref{fig:bif_asymF_asymT}(d) we can see the relevant upper and middle branches that are connected by a branch of saddle equilibria. The intermittent transition occurs because, as $F$ is increased from $-20$ mSv to $0$ mSv, the saddle moves away from the middle branch and towards the upper branch. In other words, the basin of attraction becomes larger for the middle branch and smaller for the upper branch. Therefore, near $-20$ mSv it is more likely for the noise to kick the system to the upper equilibrium, while near $0$ mSv it has become more likely to kick the system to the middle equilibrium. 
Note that the occurence of the intermittent transition is not critically dependent on the exact noise level and still occurs for other values of $\epsilon$; however, it then appears for slightly different values of $F$, since the noise strength effects the escape rates between the co-existing equilibria.

\section{Discussion}
\label{sec:discussion}

When discussing the bistability Stommel discovered in the two-box model of his seminal work, he wrote that
``one wonders
whether other quite different states of flow
are permissible in the ocean or some estuaries
and if such a system might jump into one
of these with a sufficient perturbation''.
With the introduction of only one additional box, Welander showed that ``quite different states'' could in theory exist across an inter-hemispherical ocean, such as the Atlantic Ocean \cite{WEL86}. 
In this study we have used a tailored three-box model to investigate different states in dependence of whether deep-water formation occurs at one site or two in the North Atlantic Ocean. The parameters of the model were chosen such that the differences between the boxes reflect the formation of deep water in the Nordic Seas and the Labrador Sea.  

We found that, despite its simplicity, the model captures key behavioural elements which have been observed in sophisticated ESMs. Its simplicity is of course its primary advantage. It allowed us to analyse the model and describe its possible dynamics in order to provide a clearer understanding of behaviour observed in more complex models. Given present concerns about the threat of AMOC slowdown/shutdown (for example \cite{caesar18,caesar21,boers21}), it is very important to have a sound theoretical basis of what to expect as the state of the Atlantic Ocean moves towards a critical transition. In particular, we highlight here that, instead of speaking of a single tipping \emph{event}, we expect a tipping \emph{scenario} involving multiple processes. In particular, it seems that the weakening and shutdown of convection in the Labrador Sea may act as a precursor to more drastic changes to come. 

In future work, finer detail of possible tipping scenarios could be investigated by incorporating more dynamical features into the present box model.
For example, the two equilibria on the upper branch in Fig.~\ref{fig:rahmstorf}(a) may be the result of yet another hysteresis loop, possibly due to slightly different co-existing convection patterns in the North Atlantic Ocean. Capturing this behaviour in a box model would require additional boxes in order to represent the more intricate convection patterns observed in Ref.~\cite{RAH95}.

Low-frequency variability of AMOC is a feature of the ocean dynamics that is not included in the box model. If the flow passing through boxes~$L$ and~$N$ were periodic (or periodically forced), then we would have two coupled oscillators resulting in dynamics on a torus. Therefore, there may be a direct link between the results presented in this study and so-called \emph{multifrequency tipping}, which involves tipping via a sequence of bifurcations and has previously been suggested as a possible mechanism for the dynamical features seen in Figs.~\ref{fig:genie}--\ref{fig:SCH07} \cite{keane20}.

There is another possible link between the work presented here and yet another form of tipping in the literature. When discussing the differences between the parameter values of the boxes in the model, we spoke of symmetry and asymmetry. Another perspective of the introduced asymmetry is the notion of spatial heterogeneity. This notion was recently studied in the context of tipping in Ref.~\cite{bastiaansen22}, where the heterogeneity was found to allow the existence of intermediate states with one part of the spatial domain in one state and one part in another state. The tipping of the system thus involves transitions via such intermediate states --- and is called \emph{fragmented tipping}.
Therefore, the tipping scenario of the asymmetric box model presented here could be viewed as a limiting case of fragmented tipping, where the spatial domain is discretised into only three regions (i.e. boxes).

\section*{Acknowledgements}

The project was supported by University College Cork in the framework of the \textit{SEFS New Connections Grant Award} scheme.  HD was funded by the European Research Council through the ERC-AdG project TAOC 
(project 101055096). The research of BK was supported by Royal Society Te Ap\={a}rangi Marsden Fund grant \#19-UOA-223. 

\appendix

\section{Stommel two-box model}
\label{app:stommel}

Since the Stommel ocean box model was introduced in Ref.~\cite{STO61}, it has reappeared many times in the literature with slight alterations in notation, usually with an additional freshwater forcing term; for example, see \cite{lohmann99,marotzke00,budd22}. The bifurcation diagram shown in Fig.~\ref{fig:stommel}(b) was calculated for the version in Ref.~\cite{dijkstra13}:
\begin{align}
    \label{eq:stommel}
    \dot{\Delta T} & = \eta_1 - T(1 + |q|),  \nonumber\\
    \dot{\Delta S} & = F - S(\eta_3 + |q|), \\
    q & = \Delta T - \Delta S, \nonumber
\end{align}
where the variables $\Delta T$ and $\Delta S$ represent the nondimensionalised temperature and salinity differences between the polar and equatorial boxes, respectively. The parameter $\eta_1$ represents a nondimensionalised atmospheric thermal forcing; the parameter $F$ (originally $\eta_2$ in Ref.~\cite{dijkstra13}) represents a nondimensionalised freshwater flux into the polar box; and the ratio of the thermal to freshwater surface restoration times is given by $\eta_3$. Figure~\ref{fig:stommel}(b) is for $\eta_1=3$ and $\eta_3=0.3$.

%\begin{acknowledgments}
% Put your acknowledgments here.
%\end{acknowledgments}


\bibliographystyle{ieeetr} 
\bibliography{references}

\end{document}